\documentclass[sigconf]{acmart}

\usepackage{booktabs} 

\usepackage[english]{babel}
\usepackage{moresize}
\usepackage{amsmath}
\usepackage{algorithmic}
\usepackage{balance}
\usepackage{comment}
\usepackage{paralist}
\usepackage{bm}
\usepackage{pgfplots}
\usetikzlibrary{pgfplots.dateplot}

\usepackage{flushend}
\usepackage[english]{babel}
\usepackage[latin1]{inputenc}
\usepackage{mathrsfs}
\usepackage{graphicx}

\usepackage{amssymb}
\usepackage{amsfonts}
\usepackage{url}
\usepackage{longtable}
\usepackage{rotating}
\usepackage{multirow}
\usepackage{mathrsfs}
\usepackage{subfigure}
\usepackage{enumitem}
\usepackage[linesnumbered,algoruled,boxed,lined]{algorithm2e}
\usepackage{adjustbox}
\usepackage{hyperref}

\usepackage{pgfplots}
\usetikzlibrary{pgfplots.dateplot}

\usepackage{filecontents}
\definecolor{tblue}{RGB}{31,119,180}
\definecolor{torange}{RGB}{255,127,14}
\definecolor{tgreen}{RGB}{44,160,44}
\definecolor{tred}{RGB}{214,39,40}
\definecolor{tpurple}{RGB}{148,103,189}

\newcommand{\hide}[1]{} 

\newcommand{\etal}{\textit{et al}.}

\newcommand{\ie}{\textit{i}.\textit{e}.}
\newcommand{\eg}{\textit{e}.\textit{g}.} 
\newcommand{\wrt}{\textit{w}.\textit{r}.\textit{t}} 
\newtheorem{Dfn}{Definition}

\setcopyright{acmcopyright}

\copyrightyear{2021}
\acmYear{2021}
\setcopyright{acmlicensed}\acmConference[SIGIR'21]{Proceedings of the 44th International ACM SIGIR Conference on Research and Development in Information Retrieval}{July 11--15, 2021}{Virtual Event, Canada}
\acmBooktitle{Proceedings of the 44th International ACM SIGIR Conference on Research and Development in Information Retrieval (SIGIR'21), July 11--15, 2021, Virtual Event, Canada}
\acmPrice{15.00}
\acmDOI{10.1145/3404835.3462972}
\acmISBN{978-1-4503-8037-9/21/07}

\def\model{MB-GMN}

\begin{document}
\fancyhead{}

\title{Graph Meta Network for Multi-Behavior Recommendation}

\author{Lianghao Xia}
\affiliation{South China University of Technology}
\email{cslianghao.xia@mail.scut.edu.cn}

\author{Yong Xu}
\affiliation{South China University of Technology}
\email{yxu@scut.edu.cn}

\author{Chao Huang}
\authornote{Chao Huang is the corresponding author.}
\affiliation{JD Finance America Corporation}
\email{chaohuang75@gmail.com}

\author{Peng Dai}
\affiliation{JD Finance America Corporation}
\email{peng.dai@jd.com}

\author{Liefeng Bo}
\affiliation{JD Finance America Corporation}
\email{liefeng.bo@jd.com}

\begin{abstract}

Modern recommender systems often embed users and items into low-dimensional latent representations, based on their observed interactions. In practical recommendation scenarios, users often exhibit various intents which drive them to interact with items with multiple behavior types (\eg,  click, tag-as-favorite, purchase). However, the diversity of user behaviors is ignored in most of existing approaches, which makes them difficult to capture heterogeneous relational structures across different types of interactive behaviors. Exploring multi-typed behavior patterns is of great importance to recommendation systems, yet is very challenging because of two aspects: i) The complex dependencies across different types of user-item interactions; ii) Diversity of such multi-behavior patterns may vary by users due to their personalized preference. To tackle the above challenges, we propose a Multi-Behavior recommendation framework with Graph Meta Network to incorporate the multi-behavior pattern modeling into a meta-learning paradigm. Our developed \model\ empowers the user-item interaction learning with the capability of uncovering type-dependent behavior representations, which automatically distills the behavior heterogeneity and interaction diversity for recommendations. Extensive experiments on three real-world datasets show the effectiveness of \model\ by significantly boosting the recommendation performance as compared to various state-of-the-art baselines. The source code is available at \emph{https://github.com/akaxlh/MB-GMN}.
\end{abstract}


%
%
%





\maketitle

\section{Introduction}
\label{sec:intro}

Recommender systems have played an important role in meeting users' personalized interests and alleviating the information overload for various applications, ranging from online review sites~\cite{wang2019knowledge}, location-based recommendation services~\cite{altenburger2019yelp,gao2019privacy}, and online retailing platforms~\cite{gu2020hierarchical,2019online}. To forecast user preference from observed user behavior data, many efforts have been devoted to Collaborative Filtering (CF) techniques (\eg, matrix factorization methods~\cite{koren2009matrix}). The key idea of CF paradigm is to consider user historical interactions, and make recommendations based on their potential common preferences with vectorized user/item representations~\cite{he2017neuralncf}.

With the recent advancements in deep learning, many neural network techniques have been proposed to augment collaborative filtering architecture, with the modeling of complex interactive patterns between users and items. For instance, NCF~\cite{he2017neuralncf} and DMF~\cite{xue2017deep} introduce the Multi-layer perceptron to enable the CF framework with the capability of learning non-linear feature interactions. In addition, by stacking multiple autoencoder networks, high-dimensional sparse user behavior data can be mapped into the low-dimensional dense representations under the reconstruction-based learning prototype~\cite{sedhain2015autorec,du2018collaborative}. Inspired by the recent success of graph neural networks, several studies are proposed to model the user-item interactions with graph relation encoder (\eg, Graph Convolutional Network~\cite{zhang2019star} or Graph Attention Network~\cite{wang2019kgat}).

Nevertheless, a common drawback of the above methods is that only single type of behavior is considered to characterize user-item collaborative relations, which may not comprehensively reflect the multi-dimensional users' preference in many real-world scenarios. In fact, we need to deal with more complex user-item relational structures--multi-behavior interaction scenario in which each interaction between user and item is naturally exhibited with relationship diversity~\cite{guo2019buying,xia2020knowledge}. For example, there may exist multiple relations (\eg, click, add-to-cart, add-to-favorite and purchase) between the same pair of user and item in e-commence platforms~\cite{cen2019representation}. Each individual view of behavior type could provide rich information to characterize users' preference from different dimensions. Some users might like to add their interested items into the favorite item set before placing the order, while others always make purchases once they add items into the cart. This motivates us to explore the multi-behavior interactive patterns in modeling users' preference that yields more accurate recommendations.

While it is desirable to make use of other types of behaviors in assisting the recommendation task, it is non-trivial to encode such multi-behavioral collaborative relations. Towards this end, the technical challenges mainly come from two perspectives:\\\vspace{-0.12in}

\noindent \textbf{Modeling with Behavior Heterogeneity}. It is challenging to learn a comprehensive representation which can comprehensively reflect preference patterns of user-item interaction from different behavior types. Different behavior type view can present complementary knowledge for learning users' interests and interweave with each other in complex ways. Furthermore, each type of user behavior may exhibit pairwise dependencies or even high-level cross-type behavior correlations. Hence, in order to learn meaningful multi-behavior representations from type-specific, inter-type behavior relations, an effective cross-type interactive feature learning model with behavior pattern fusion is a necessity in solving multi-behavior recommendation problem. \\\vspace{-0.1in}

\noindent \textbf{Learning with Interaction Diversity}. The dependencies across different types of user-item interactions are complex since user's type-specific behaviors are likely to be dependent because of various factors~\cite{wen2019leveraging}. In practical scenarios, users often exhibit their different correlated behavioral patterns due to their personal specialty and habits, \eg, some people prefer to add items into their favorite lists only if they will buy them with high probabilities. In contrast, some one may like to add many different items into his/her favorite list but only makes sporadic purchases. The inter-dependencies between multiple types of behaviors vary by users. Although a handful of studies attempt to learn users' interests from their different types of user behaviors~\cite{gao2019neural,guo2019buying,jin2020multi}, they merely consider the correlations between different types of user-item interactions under the same representation space, which can hardly capture personalized characteristics from users' multi-behavior interactions. Therefore, a sufficiently generalized knowledge transfer framework is required, to distill the comprehensive and diverse multi-behavior patterns of users.

Hence, to address the above challenges, in this paper, we propose a new \underline{M}ulti-\underline{B}ehavior recommendation framework with \underline{G}raph \underline{M}eta \underline{N}etwork (MB-GMN). The goal of \model\ is to build a customized meta-learning paradigm upon the multi-behavior relation learning. Specifically, we first propose a meta-knowledge learner for encoding behavior heterogeneity with the consideration of personalized user-specific interactions. After that, a meta graph neural network is developed to capture diverse multi-behavior patterns with high-order connectivity over the user-item interaction graph. With the graph-structured embedding propagation paradigm as our multi-behavior pattern modeling component, the behavior heterogeneity and diversity can be jointly preserved in our learned latent user/item representations. Finally, we introduce a meta prediction network to capture the complex cross-type behavior dependency under a multi-task learning framework. By doing so, multi-typed user behaviors are not only used to modulate the model parameters of graph neural models, but also guide the prediction phase with the injection of supervising signals.


In summary, our major contributions are three-fold:
\begin{itemize}[leftmargin=*]

\item To the best of our knowledge, we are the first to study the multi-behavior recommendation problem with the joint modeling behavior heterogeneity and interaction diversity. 

\item We develop a new framework, \model\ which incorporates a multi-behavior pattern modeling component with the meta-learning paradigm. The multi-behavioral relations are distilled to customize the graph neural network for capturing the personalized high-order collaborative effects. In addition, a meta prediction network is introduced to supercharge \model\ with the learning of cross-type behavior dependency.

\item We demonstrate the effectiveness of \model\ on three real-world datasets to show that our framework advances the recommendation performance as compared to baselines from various research lines. Additionally, we further perform the ablation study to better understand the model design of \model.

\end{itemize}

\section{Preliminary}
\label{sec:model}





We let $U$ and $V$ represent the set of users and items, respectively, \ie, $\mathcal{U}=\{u_1,...,u_i,...,u_I\}$ and $\mathcal{V}=\{v_1,...,v_j,...,v_J\}$, where $I$ (indexed by $i$) and $J$ (indexed by $j$) represent the numbers of users and items. With the consideration of multiplex user-item interactions, we define a multi-behavior interaction tensor as below:

\begin{Dfn}
\textbf{Multi-Behavior Interaction Tensor} $\textbf{X}$. We define a three-way tensor $\textbf{X}\in \mathbb{R}^{I\times J\times K}$ to reflect the multi-typed interaction (\eg, click, tag-as-favorite, purchase) between user ($u_i \in \mathcal{U}$) and item ($v_j \in \mathcal{V}$). Here, $K$ denotes the number of behavior types. Specifically, the individual element $x_{i,j}^k \in \textbf{X}$ is set as 1 if user $u_i$ interacts with item $v_j$ under the $k$-th behavior type, and $x_{i,j}^k=0$ otherwise.
\end{Dfn}


In the multi-behavior recommendation scenario, one type of user-item interaction will be treated as \emph{target behavior} (\eg, purchase). Other behaviors are referred to as \emph{context behaviors} (\eg, click, tag-as-favorite, add-to-cart) for providing insightful knowledge in assisting the target behavior prediction. Based on the aforementioned definitions, we formally state our studied problem as:
\begin{itemize}[leftmargin=*]
    \item \textbf{Input}: the observed multi-behavior interaction tensor $\textbf{X} \in \mathbb{R}^{I\times J\times K}$ between users $\mathcal{U}$ and items $\mathcal{V}$ across $K$ behavior types. \\\vspace{-0.1in}
    \item \textbf{Output}: the predictive function which estimates the likelihood of user $u_i$ adopts the item $v_j$ with the target behavior type.
\end{itemize}

\section{Methodology}
\label{sec:solution}

\begin{figure*}
    \centering
    \includegraphics[width=\textwidth]{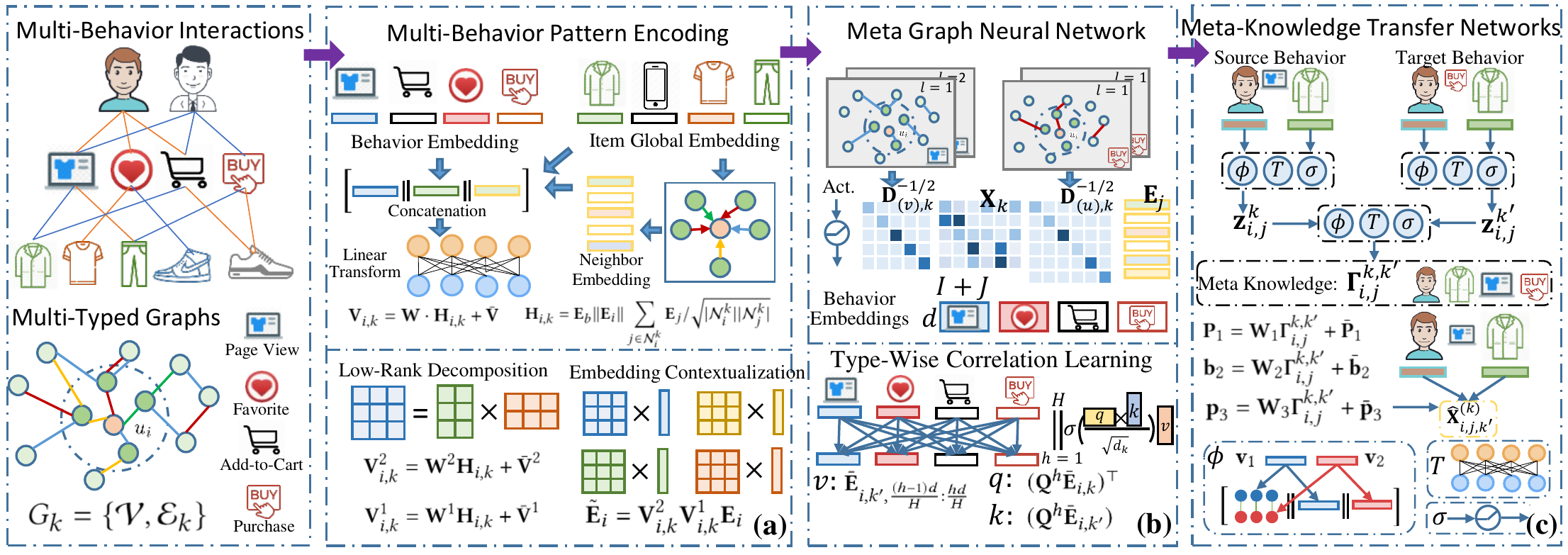}
    \vspace{-0.15in}
    \caption{The model architecture of \model. (a) Multi-behavior pattern modeling with meta-knowledge learner for behavior heterogeneity; (b) Meta graph neural network which preserves the behavior semantics with high-order connectivity; (c) Meta-knowledge transfer networks that customize the parameter of prediction layer to capture cross-type behavior dependency.}
    \label{fig:framework}
\end{figure*}


We now elaborate the proposed \model, the model flow of which is illustrated in Figure~\ref{fig:framework}. \model\ is composed of two key components: i) multi-behavior pattern encoding, a meta-knowledge learner that captures the personalized multi-behavior characteristics; ii) cross-type behavior dependency modeling, a transfer learning paradigm which learns a well-customized prediction network by transferring knowledge across different behavior types.


\subsection{Multi-Behavior Pattern Modeling}
Unlike previous multi-behavior recommendation models~\cite{jin2020multi,gao2019neural,xia2020multiplex} which parameterize multi-typed interaction patterns of different individuals into the same representation space, we aim to capture the personalized multi-behavior characteristics corresponding to diverse user preferences. Towards this end, we propose an integrative neural architecture with a meta-knowledge learner and meta graph neural network. This new framework jointly captures the user-item relation heterogeneity under multi-typed behavior context, as well as the mutual relationships between different behaviors, with respect to their user-specific patterns.


\subsubsection{\bf Meta-knowledge Learner for Behavior Heterogeneity}
In this component, we aim to learn the meta-knowledge from user-specific multi-behavior interactive patterns over different items. We then utilize the learned meta-knowledge to generate the transformation weights for injecting the type-specific behaviour semantics into the initial embeddings.
Specifically, given the initialized ID-corresponding embeddings $\textbf{E}_i\in\mathbb{R}^d$ and $\textbf{E}_j\in\mathbb{R}^d$ for user $u_i$ and item $v_j$, respectively, we exert the meta network to learn a personalized context projection layer with latent transformations:
\begin{align}
    \textbf{V}_{i,k}=\textbf{W}\cdot\textbf{H}_{i,k}+\bar{\textbf{V}};~~~~
    \textbf{H}_{i,k}=\textbf{E}_b \| \textbf{E}_i \| \sum_{j\in\mathcal{N}_{i}^k} {\textbf{E}_j} / {\sqrt{|\mathcal{N}_i^k| |\mathcal{N}_j^k|}}
\end{align}
\noindent where $\|$ denotes the concatenation operation. $\textbf{V}_{i,k}\in\mathbb{R}^{d\times d}$ is the learned customized parametric matrix for user $u_i$, which injects the type-specific behavior context into personalized user representation. $\textbf{W}\in\mathbb{R}^{d\times d\times d}$ and $\bar{\textbf{V}}\in\mathbb{R}^{d\times d}$ represent the transformation parameters of the meta network. $\textbf{H}_{i,k}\in\mathbb{R}^d$ is the context-aware embedding which preserves type-specific user interaction behavior patterns. Here, we adopt $\sqrt{|\mathcal{N}_i^k| |\mathcal{N}_j^k|}$ as the normalization term for embedding aggregation. In general, this proposed meta-knowledge learner takes the user-specific behavior characteristics as input, \ie, $\textbf{E}_i$, $\textbf{E}_b$, as well as personalized interactions over his/her connected items $\sum_{j\in\mathcal{N}_{i}^k} {\textbf{E}_j}$ under the $k$-th behavior type, where $\mathcal{N}_i^k$ denotes the set of interacted items of user $u_i$. The model output is the parametric matrix $\textbf{V}_{i,k}\in\mathbb{R}^{d\times d}$ which is used to generate weights of graph neural network for handling behavior heterogeneity.



\subsubsection{\bf Low-rank Transformation Decomposition} While we can express the customized context representations for user-specific multi-behaviors, it is challenging to extract such meta-knowledge in an efficient way. To tackle this challenge, we enhance our meta-knowledge learner with the low-rank decomposition technique for transformation operation. In particular, the learning phase over the meta-knowledge transformation $\textbf{V}_{i,k}\in\mathbb{R}^{d\times d}$ can be decomposed into two low-rank projections as below:
\begin{align}
    \tilde{\textbf{E}}_i=\textbf{V}_{i,k}^2 \textbf{V}_{i,k}^1 \textbf{E}_i;~
    \textbf{V}_{i,k}^1=\textbf{W}^1\textbf{H}_{i,k}+\bar{\textbf{V}}^1;~
    \textbf{V}_{i,k}^2=\textbf{W}^2\textbf{H}_{i,k}+\bar{\textbf{V}}^2;
\end{align}

\noindent where $\tilde{\textbf{E}}_i\in\mathbb{R}^d$ is the contextualized user embedding of user $u_i$, and $\textbf{V}_{i,k}^1\in\mathbb{R}^{d'\times d}, \textbf{V}_{i,k}^2\in\mathbb{R}^{d\times d'}$ are the decomposed transformation matrices, with $\textbf{W}^1,\textbf{W}^2,\bar{\textbf{V}}^1,\bar{\textbf{V}}^2$ as trainable parameters. Note that $d'$ is the low-rank state dimensionality which is much smaller than the regular embedding dimension size $d$. Hence, the transformation integration $\textbf{V}_{i,k}=\textbf{V}_{i,k}^2\textbf{V}_{i,k}^1$ is restrict to be low-rank. For the item side, the corresponding embeddings $\textbf{E}_j$ is contextualized to yield $\tilde{\textbf{E}}_j$ under the same meta-knowledge learning framework.

\subsection{Meta Graph Neural Network}
To capture the diverse behavior patterns of different users, we propose a meta graph neural network to enhance latent representation learning, with the preservation of behavior-aware user-item collaborative effects. There are three key sub-modules in the meta graph neural network, \ie, i) Behavior semantic encoding; ii) Behavior mutual dependency learning; iii) High-order multi-behavioral context aggregation. We will describe the technical details of each module in the following subsections.

\subsubsection{\bf Behavior Semantic Encoding}
In our recommendation scenario with multi-behavioral relations, each individual type of interaction behavior has its own characteristics. Taking the e-commence platform
as the example, user's page view behavior is more likely to happen than purchase. Moreover, add-to-cart and purchase behavior may co-occur with high probability. In such cases, we propose to capture the personalized behavior semantic signals during the multi-behavior pattern modeling. Given the learned the meta-knowledge $\tilde{\textbf{E}}_i$ of each user $u_i$ and $\tilde{\textbf{E}}_j$ of item $v_j$, we design a behavior-aware message passing strategy to capture user-item graph structure $G_k=\{\mathcal{V}, \mathcal{E}_k\}$, where $\mathcal{V}$ denotes the set of nodes (\ie, users and items), and $\mathcal{E}_k$ represents the set of interaction edges among $\mathcal{V}$ under the $k$-th behavior type. The behavior context-aware embeddings of user $u_i$ and item $v_j$ are generated as follows:
\begin{align}
    \bar{\textbf{E}}_{i,k} = \textbf{E}_i + \sigma(\sum_{j\in\mathcal{N}_i^k} \alpha_{i,j,k} \textbf{E}_j);~~~~~~
    \bar{\textbf{E}}_{j,k} = \textbf{E}_j + \sigma(\sum_{i\in\mathcal{N}_j^k} \alpha_{i,j,k} \textbf{E}_i)
\end{align}
\noindent where $\bar{\textbf{E}}_{i,k},\bar{\textbf{E}}_{j,k} \in \mathbb{R}^d$ are the smoothed embeddings for user $u_i$ and item $v_j$ under the behavior type of $k$, respectively. We define $\alpha_{i,j,k}$ as the normalization factor for $u_i$-$v_j$ interaction with the $k$-th behavior type, which is formally given as $\alpha_{i,j,k} = 1 / \sqrt{|\mathcal{N}_i^k||\mathcal{N}_j^k|}$. We summarize the above embedding propagation procedure over user-item interaction $G_k$ with the matrix format as:
\begin{align}
\bar{\textbf{E}}_{(u),k}=\textbf{E}_{(u)}+ \sigma(\textbf{D}_{(u),k}^{-1/2}\cdot \textbf{X}_k\cdot\textbf{D}_{(v),k}^{-1/2} \textbf{E}_{(v)})
\end{align}
\noindent where $\textbf{X}_k\in\mathbb{R}^{N\times M}$ denotes the adjacent matrix of graph $G_k$ under the $k$-th behavior type. $\textbf{D}_{(u),k}^{-1/2}$ and $\textbf{D}_{(v),k}^{-1/2}$ are the diagonal degree matrices for users and items. Here, $\textbf{E}_{(u)}\in\mathbb{R}^{N\times d}, \textbf{E}_{(v)}\in\mathbb{R}^{M\times d}$ corresponds to the user and item embedding tables, respectively. In \model, we further employ a residual unit to prevent the gradient vanishing issue. Inspired by the convolution optimization scheme for collaborative filtering in LightGCN~\cite{he2020lightgcn}, \model\ is not equipped with linear transformations during the message passing. Therefore, with the learned meta-knowledge of behavior heterogeneity, \model\ endows the graph-structured information propagation with the interaction modeling in a customized manner under personalized behavior context of individual user.

\subsubsection{\bf Behavior Relation Learning}
In practice, different types of interaction behavior are correlated in a complex way, which brings new challenge for modeling multi-behavioral
relationships. To encode the underlying relations across different type of behaviors, we design a multi-behavior relation encoding function, to refine the propagated embeddings by injecting the mutual relational information between different behavior types. The relation encoding function is built upon the attention network:
\begin{align}
    \bar{\textbf{E}}_{i,K+1}=\sum_{k=1}^K{\textbf{H}}_{i,k};~~~~
    {\textbf{H}}_{i,k}= \mathop{\Bigm|\Bigm|}\limits_{h=1}^H \sum_{k'=1}^K\beta_{k,k'}^h\cdot\bar{\textbf{E}}_{i,k',\frac{(h-1)d}{H}: \frac{hd}{H}};\nonumber\\
    \beta_{k,k'}^h=\frac{\exp{\hat{\beta}_{k,k'}^h}}{\sum_{k'=1}^K\exp{\hat{\beta}_{k,k'}^h}}; ~~~~~~~
    \hat{\beta}_{k,k'}^h = \frac{(\textbf{Q}^h\bar{\textbf{E}}_{i,k})^\top(\textbf{Q}^h\bar{\textbf{E}}_{i,k'})}{\sqrt{d/H}}
\end{align}
where $\tilde{\textbf{E}}_{i,K+1}\in\mathbb{R}^d$ denotes the global user embedding that takes all behavior types into account. ${\textbf{H}}_{i,k}\in\mathbb{R}^d$ is an intermediate representation which encodes the behavior-specific features refined by other behavioral data. ${\textbf{H}}_{i,k}$ is calculated using an $H$-head dot-product attention mechanism. Additionally, $\beta_{k,k'}^h\in\mathbb{R}$ is the $h$-th quantitative weight for behavior type $k$ and $k'$, subscript $\frac{(h-1)d}{H}: \frac{hd}{H}$ denotes the $h$-th slice of the vector (each slice is of size $d/H$), and $\textbf{Q}^h\in\mathbb{R}^{\frac{d}{H}\times d}$ is the $h$-head parametric transformation. To alleviate the overfitting effect and improve the training efficiency, we optimize our attentive encoding function from the following perspectives: 1) we slice the $d$-dimensional $\bar{\textbf{E}}_{i,k'}$ to $H$ parts as feature vectors corresponding to $H$ heads, without the heavy value transformation; 2) we enable the matrix transformation sharing among the query and key dimensions with the projection parameter $\textbf{Q}^h$.

\subsubsection{\bf High-order Multi-Behavioral Context Aggregation}
To capture the high-order connectivity of the interaction graph, \model\ integrates the behavior semantic encoding with the behavior-wise relational structure learning under the high-order embedding propagation paradigm. In specific, we inject the high-order connectivity into the user embedding $u_i$ as:
$$ \bar{\textbf{E}}_{k}^{(l+1)}=\left\{
\begin{aligned}
\text{Graph-Conv}(\bar{\textbf{E}}_{k}^{(l)}),~~~~ k=1,2,...,K \\
\text{Att}(\bar{\textbf{E}}_{1}^{(l+1)}, ..., \bar{\textbf{E}}_{K}^{(l+1)}),~~ k=K+1
\end{aligned}
\right.
$$
\noindent where $\text{Graph-Conv}(\cdot)$ denotes the behavior semantic encoder, and $\text{Att}(\cdot)$ represents the function for behavior dependency learning. By stacking the graph encoders with $L$ times, \model\ is aware of the $L$-hop graph-based connection structure for each node. To aggregate the cross-order message embeddings, we integrate the order-specific embeddings and yield the final user and item representations using the operation: $\hat{\textbf{E}}_k=\sum_{l=0}^L\bar{\textbf{E}}^l_k$ for $k=1,2,...,K+1$.


\subsection{Meta-Knowledge Transfer Networks}
With the design of our meta graph neural network, we realize the information propagation and aggregation between users and items, so as to characterize the personalized multi-behavior patterns vary by users. Now, we move on to the prediction of different types of user-item interactions, based on a meta learning paradigm for customized knowledge transfer. From the meta graph neural network, we could obtain the $(K+1)$ aggregated behavior representations, \ie, type-specific behavioral patterns are characterized by $K$ different embeddings $\hat{\textbf{E}}_k$, and a summarized general interaction (without differentiating behavior types) $\hat{\textbf{E}}_{K+1}$ which preserves the general user interest. In the prediction phase, how to effectively capture the cross-type behavior dependency and transfer the influential knowledge in assisting the forecasting task on the target behavior, remains a significant challenge. 

\subsubsection{\bf Meta-knowledge Learner for Behavior Dependency} In \model, we propose a meta learning scheme to generate prediction sub-networks with respect to the target user-item pair ($u_i$, $v_j$) and behavior type. During our target behavior (\eg, purchase) prediction phase, context behavior information (\eg, click, tag-as-favorite, add-to-cart) not only provides the insightful external knowledge, but also serve as the supervising signals for model optimization. Given user-item pair $u_i$ and $v_j$, our meta prediction network first learns the underlying meta knowledge customized to $u_i$ and $v_j$ under the target behavior type of $k'$ (other context behavior types are indexed by $k$). This process is presented as follows:
\begin{align}
    \mathbf{\Gamma}_{i,j}^{k,k'}=\sigma(\textbf{W}_1\cdot\phi(\textbf{Z}_{i,j}^k, \textbf{Z}_{i,j}^{k'}));~~~~ \textbf{Z}_{i,j}^k=\sigma(\textbf{W}_2\cdot\phi(\hat{\textbf{E}}_{i,k}, \hat{\textbf{E}}_{j,k}))
\end{align}
\noindent where $\mathbf{\Gamma}_{i,j}^{k,k'}\in\mathbb{R}^d$ is the encoded meta knowledge of $u_i$ and $v_j$ with the relation between the target behavior $k'$ and context behaviors $k$ ($k\in (K+1)$, $k'\in K$). $\textbf{Z}_{i,j}^k,\textbf{Z}_{i,j}^{k'}\in\mathbb{R}^d$ are the projected representations of $u_i$ and $v_j$ with the relation between the $k'$-th and the $k$-th behavior type. $\sigma(\cdot)$ is the activation function and $\phi(\cdot)$ is the encoding function to capture the dependent interactive relations between two embeddings. To make use of both the binary relations and the unary bias, \model\ proposes to employ a $\phi(\cdot)$ function defined as: $\phi(\textbf{v}_1, \textbf{v}_2)=\textbf{v}_1\circ\textbf{v}_2 \| \textbf{v}_1 \| \textbf{v}_2$, where $\circ$ denotes element-wise multiplication and $\|$ denotes concatenation operation.

\subsubsection{\bf Knowledge Transfer Learning Framework}
With the meta-knowledge representation $\mathbf{\Gamma}_{i,j}^{k,k'}$, \model\ generates three parametric variables for our prediction network as follows:
\begin{align}
    \textbf{P}_1=\textbf{W}_1 \mathbf{\Gamma}_{i,j}^{k,k'}+\bar{\textbf{P}}_1;~~
    \textbf{b}_2=\textbf{W}_2 \mathbf{\Gamma}_{i,j}^{k,k'}+\bar{\textbf{b}}_2;~~
    \textbf{p}_3=\textbf{W}_3 \mathbf{\Gamma}_{i,j}^{k,k'}+\bar{\textbf{p}}_3
\end{align}
where $\textbf{P}_1\in\mathbb{R}^{d\times 3d}, \textbf{b}_2, \textbf{p}_3 \in\mathbb{R}^d$ are the learned parameters for $u_i$, $v_j$, source task $k$ and target task $k'$. $\textbf{W}_*$ are the parameters of the meta network, and $\bar{\textbf{P}}_1, \bar{\textbf{b}}_2, \bar{\textbf{p}}_3$ are the bias parameters for $\textbf{P}_1, \textbf{b}_2$ and $\textbf{p}_3$. Finally, \model\ makes predictions on $u_i$, $v_j$ under the target behavior $k'$, using source data of context behavior $k$, by:
\begin{align}
    \hat{\textbf{X}}_{i,j,k'}^{(k)}=\mathbf{\eta}^\top\textbf{p}_3;~~~~
    \mathbf{\eta} = \sigma(\textbf{P}_1 \cdot \phi(\hat{\textbf{E}}_{i,k}, \hat{\textbf{E}}_{j,k}) + \textbf{b}_2)
\end{align}
where $\hat{\textbf{X}}_{i,j,k'}^{(k)}\in\mathbb{R}$ is the predicted likelihood of $u_i$ interacting with $v_j$ by behavior $k'$, given the interaction data $k$ as features. $\mathbf{\eta}\in\mathbb{R}^d$ is an intermediate feature vector. For the most concerned target behavior $k'$ (\eg~ purchase), we use $\hat{\textbf{X}}_{i,j,k'}^{(K+1)}$ as the final prediction.

\subsection{Multi-task Learning Framework}
We train \model\ by optimizing the prediction objective for each pair of source and target behavior. For user $u_i$ and target behavior $k'$, \model\ samples $S$ positive interactions and $S$ non-interacted items as the negative samples. During the training phase, we adopt the Adam algorithm to optimize the following defined objective:
\begin{align}
    \mathcal{L}=\sum_{i=1}^N\sum_{k=1}^{K+1}\sum_{k'=1}^K\sum_{s=1}^S \max(0, 1-\hat{\textbf{X}}_{i,p_s,k'}^{(k)}+\hat{\textbf{X}}_{i,n_s,k'}^{(k)}) + \lambda \|\mathbf{\Theta}\|_\text{F}^2
\end{align}
where $k$ refers to the source behavior and $k'$ refers to the target behavior, $p_s$ and $n_s$ are the $s$-th positive and negative samples, respectively. The first term is the marginal pair-wise loss, and the last term denotes the weight-decay regularization with weight $\lambda$.

\subsubsection{\bf Model Complexity Analysis}
The time cost of \model\ mainly comes from the following aspects. i) In the multi-behavior pattern modeling, \model\ adopts lightweight graph convolutional architecture which costs only $O(L\times K\times d \times |\mathcal{E}|)$ across $L$ layers, $K$ behavior types, $d$ latent factors and $|\mathcal{E}|$ edges. The behavior relation learning costs extra $O(L\times K\times d\times(K+d)\times (N+M))$. As $O(d\times |\mathcal{E}|)$ is comparable with $O((K+d)\times(N+M))$ in our case, the complexity doesn't increase. The contextualization meta network costs $O(K\times (N+M)\times d^2\times d')$ for parameter learning, and $O(K\times (N+M)\times d\times d')$ for embedding transformation, which is very close to the complexity of GCN networks. The prediction meta networks costs minor $O(S\times d^2)$ computations for each user. In conclusion, our \model\ could achieve comparable model time complexity with some graph convolution-based models.


\section{Evaluation}
\label{sec:eval}

This section performs experiments on different datasets to evaluate the performance of our \emph{\model} framework by making comparison with various state-of-the-art recommendation techniques. In particular, we aim to answer the following research questions:

\begin{itemize}[leftmargin=*]

\item \textbf{RQ1}: How does \emph{\model} perform when competing with various recommendation baselines on different datasets? \\\vspace{-0.1in}

\item \textbf{RQ2}: How do the designed sub-modules in our \emph{\model} framework affect the recommendation performance? \\\vspace{-0.1in}

\item \textbf{RQ3}: How does \emph{\model} work for making recommendation to users with the consideration of different types of behaviors? \\\vspace{-0.1in}

\item \textbf{RQ4}: How does \emph{\model} perform when handling behavior data with different interaction sparsity degrees? \\\vspace{-0.1in}


\item \textbf{RQ5}: How do different configurations of key hyperparameters impact the performance of \emph{\model} framework? \\\vspace{-0.1in}

\item \textbf{RQ6}: How is the model explainability of our \emph{\model}? What relational patterns between different types of behaviors could \emph{\model} capture in assisting the final recommendation task?

\end{itemize}

\begin{table}[t]
    \caption{Statistics of experimented datasets}
\vspace{-0.2in}
    \label{tab:data}
    \centering
    \footnotesize
	\setlength{\tabcolsep}{0.6mm}
    \begin{tabular}{ccccc}
        \midrule
        Dataset&User \#&Item \#&Interaction \#& Interactive Behavior Type\\
        \hline
        Taobao-Data& 147894 & 99037 & 7658926 & \{Page View, Favorite, Cart, Purchase\}\\
        Beibei-Data& 21716 & 7977 & 3338068 &\{Page View, Cart, Purchase\}\\
        IJCAI-Contest& 423423 & 874328 & 36203512 &\{Page View, Favorite, Cart, Purchase\}\\
        \hline
    \end{tabular}
\vspace{-0.1in}
\end{table}

\subsection{Experiment Settings}

\subsubsection{\bf Dataset}
\label{sec:data}

To evaluate the effectiveness of our model, we conduct experiments on three datasets from real-world platforms. We describe the data details as below. \\\vspace{-0.12in}

\noindent \textbf{Taobao Data}. There are four types of user-item relations contained in this data from Taobao--one of the largest e-commerce platform in China, \ie, \textit{page view}, \textit{add-to-cart}, \textit{tag-as-favorite} and \textit{purchase}. \ \\\vspace{-0.12in}

\noindent \textbf{Beibei Data}. This data is collected from Beibei which is one of the largest infant product online retailing site in China. It involves three types of user-item interactions, including \emph{page view}, \emph{add-to-cart} and \emph{purchase}, to generate the multi-behavior interaction tensor $\textbf{X}$. \\\vspace{-0.12in}

\noindent \textbf{IJCAI Data}. This data is released by IJCAI competition for user activity modeling from an online business-to-consumer e-commerce platform. Four types of behaviors are included in this dataset, \ie, \emph{page view}, \emph{add-to-cart}, \emph{tag-as-favorite} and \emph{purchasing}. \\\vspace{-0.12in}

In those three datasets, we regard users' purchases as the target behaviors, because purchases are good indicators for Gross Merchandise Volume which indicates the total volume of sales in real-life online retailing sites~\cite{guo2019buying,wu2018turning}. We further summarize the detailed information of each experimented dataset in Table~\ref{tab:data}.

\subsubsection{\bf Evaluation Protocols}
For performance evaluations, we use two representative metrics: \textit{Normalized Discounted Cumulative Gain (NDCG@$N$)} and \textit{Hit Ratio (HR@$N$)} which have been widely adopted in top-N recommendation tasks~\cite{wang2019neural,bai2019ctrec}. Following the same experimental settings in~\cite{he2016fast,he2017neuralncf}, the leave-one-out evaluation is leveraged for training and test set partition. We generate the test data set by including the last interactive item and consider the rest of data for training. For efficient and fair model evaluation, we pair each positive item instance with 99 randomly sampled non-interactive items for each user, which shares similar settings in~\cite{sun2019bert4rec,kang2018self}.

\subsubsection{\bf Methods for Comparison}
\label{sec:baseline}
To comprehensively evaluate the performance, we compare our \emph{\model} against various baselines from different research lines, which are elaborated as below:

\noindent \textbf{Conventional Matrix Factorization Approach}:
\begin{itemize}[leftmargin=*]
\item \textbf{BiasMF}~\cite{koren2009matrix}: this approach considers the bias information from users and items using the matrix factorization model.

\end{itemize}

\noindent \textbf{Neural Collaborative Filtering Methods}:
\begin{itemize}[leftmargin=*]
\item \textbf{DMF}~\cite{xue2017deep}: this model considers the explicit interaction and non-preference implicit feedback into a deep matrix factorization model for user/item embedding mapping.
\item \textbf{NCF}~\cite{he2017neuralncf}: we consider three variants of NCF with the configuration of different interaction encoders: \ie, concatenated element-wise-product branch (\ie, NCF-N), Multilayer perceptron (\ie, NCF-M), and the fixed element-wise product (\ie, NCF-G).
\end{itemize}

\noindent \textbf{Autoencoder-based Collaborative Filtering Models}:
\begin{itemize}[leftmargin=*]
\item \textbf{CDAE}~\cite{wu2016collaborative}: it applies the denoising autoencoders to train a neural network with the data reconstruction objective in an adaptive way.
\item \textbf{AutoRec}~\cite{sedhain2015autorec}: it stacks multi-layer autoencoder to map user-item interactions into latent low-dimensional representations.
\end{itemize}

\noindent \textbf{Neural Auto-regressive Recommendation Models}:
\begin{itemize}[leftmargin=*]
\item \textbf{NADE}~\cite{zheng2016neural}: it is a feed-forward autoregressive architecture for collaborative filtering with the parameter sharing technique.
\item \textbf{CF-UIcA}~\cite{du2018collaborative}: it explicitly captures the relations between items and users with a neural co-autoregressive model.
\end{itemize}

\noindent \textbf{Graph Neural Networks for Recommender Systems}:
\begin{itemize}[leftmargin=*]
\item \textbf{ST-GCN}~\cite{zhang2019star}: this method is a convolution-based graph neural model which generates user embeddings based on an encoder-decoder framework.
\item \textbf{NGCF}~\cite{wang2019neural}: it is a message passing architecture for information aggregation over the user-item interaction graph, to exploit high-order connection relationships.
\end{itemize}

\noindent \textbf{Recommendation with Multi-Behavior Relations}.
\begin{itemize}[leftmargin=*]
\item \textbf{NMTR}~\cite{gao2019neural}: it is a multi-task recommendation framework for performing the cascade prediction of different types of behaviors.
\item \textbf{DIPN}~\cite{guo2019buying}: it aims to predict the purchase intent of users by combing different user interactive patterns based on attention.
\item \textbf{NGCF$_M$}~\cite{wang2019neural}: we enhance the NGCF model to inject the multi-behavioral graph relations under a graph neural network.
\item \textbf{MATN}~\cite{xia2020multiplex}: it differentiates the relations between user and item with the integration of the attention network and memory units.
\item \textbf{MBGCN}~\cite{jin2020multi}: it designs a unified graph to represent the multi-behavior of users and uses graph convolutional network to perform behavior-aware embedding propagation.
\end{itemize}


\subsubsection{\bf Parameter Settings}
We implement the proposed \model\ model using TensorFlow. The parameters are learned using the Adam optimizer, with the learning rate $1e^{-3}$ and decay rate 0.96. The regularization weight $\lambda$ is selected from \{0.05, 0.01, 0.005, 0.001, 0.0005, 0.0001\}. The training batch size is selected from \{32, 128, 256, 512\}. By default, \model\ is configured with the hidden state dimensionality of 16, low-rank dimensionality of 4, 2 attention heads. The number of message passing layers in our meta graph neural network is searched from \{1,2,3\}. The experiments for most of neural network baselines are performed based on their released codes or presented settings. For NCF and NMTR, we sample the positive and negative instances with the range of 1 : 1 to 1 : 4 to fit their model optimization with point-wise loss.





\subsection{Performance Validation (RQ1)}
We evaluate the performance of all baseline methods in predicting the target user interactive behavior on different datasets, and summarize the following observations:

Table~\ref{tab:target_behavior} shows the performance of top-$N$ item recommendation under the target behavior type and ``Imp'' indicates the relatively improvement ratio between our \emph{\model} and each baseline. From the presented results, we observe that the recommendation performance could be obviously improved by \emph{\model}. Such performance gap can be attributed to the effective modeling of personalized multi-behavior patterns, which results in customized user/item representations under a meta learning paradigm.

We can also observe that \emph{\model} consistently obtains better performance than several baselines which models multi-typed user behaviors for recommendations. For example, the evaluation results shed light on the limitation of NMTR method, by merely modeling the singular dimensional cascading correlations between multi-type interactions. The dual-stage attention mechanisms (DIPN and MATN) which aggregate different types of behavioral patterns through weighted summation, cannot comprehensively capture the complex inter-dependencies across different types of user behavior. Additionally, the large performance gap between \emph{\model} and compared multi-behavior graph neural frameworks (\ie, NGCF$_M$ \& MBGCN), justifies the advantage of our method for performing customized multi-behavior dependency modeling. In contrast, these GNN-based baselines aggregate the behavior-aware patterns under a parameter sharing message passing architecture, which overlook the unique characteristics of user behavior patterns. \\\vspace{-0.1in}

Among various compared methods, we can observe that the injection of multi-behavior
information into the recommendation frameworks (\ie, NMTR, DIPN, NGCF$_M$, MATN, MBGCN) could boost the performance, as compared to other baselines which do not differentiate user-item interactions. This observation confirms the benefits of exploring multi-behavior patterns for modeling collaborative effects in recommendation. Furthermore, GNN-based models are superior to other baselines, which suggests that high-order relations pertinent to users and items are derived with stacking message passing layers. \\\vspace{-0.1in}

We also evaluate the performance of \emph{\model} and several representative baselines while varying the value of $N$ in terms of HR and NDCG. Table~\ref{tab:vary_k} shows the evaluation results on Beibei dataset. We could observe that \emph{\model} always achieves the best performance with different top-$N$ positions. This observation indicates the consistent superiority of our developed framework as compared to other baselines, in giving correctly interacted items with higher probability.

\begin{table*}[t]
\caption{Overall model performance on Beibei, Taobao and IJCAI data, with the metrics of \textit{HR@$N$} and \textit{NDCG@$N$} ($N=10$).}
\vspace{-0.1in}
\centering
\footnotesize
\setlength{\tabcolsep}{1mm}
\begin{tabular}{|c|c|c|c|c|c|c|c|c|c|c|c|c|c|c|c|c|c|c|}
\hline
Data & Metric & BiasMF & ~DMF~ & NCF-M & NCF-G & NCF-N & AutoRec & ~CDAE~ & NADE & CF-UIcA & ST-GCN & NGCF & NMTR & DIPN & NGCF$_M$ & MBGCN & MATN & \emph{\model}\\
\hline
\multirow{4}{*}{Beibei}
&HR & 0.588 & 0.597 & 0.595 & 0.594 & 0.601 & 0.607 & 0.608 & 0.608 & 0.610 & 0.609 & 0.611 & 0.613 & 0.631 & 0.634 & 0.642 & 0.626 & \textbf{0.691}\\
\cline{2-19}
&Imprv & 17.52\% & 15.75\% & 16.13\% & 16.33\% & 14.98\% & 13.84\% & 13.65\% & 13.65\% & 13.28\% & 13.46\% & 13.09\% & 12.72\% & 9.51\% & 8.99\% & 7.63\% & 10.38\% & --\\
\cline{2-19}
&NDCG & 0.333 & 0.336 & 0.332 & 0.338 & 0.336 & 0.341 & 0.341 & 0.343 & 0.346 & 0.343 & 0.375 & 0.349 & 0.394 & 0.372 & 0.376 & 0.385 & \textbf{0.410}\\
\cline{2-19}
&Imprv & 23.12\% & 22.02\% & 23.49\% & 21.30\% & 22.02\% & 20.23\% & 20.23\% & 19.53\% & 18.50\% & 19.53\% & 9.33\% & 17.48\% & 4.06\% & 10.22\% & 9.04\% & 6.49\% & --\\
\hline
\multirow{4}{*}{Taobao}
&HR & 0.262 & 0.305 & 0.319 & 0.290 & 0.325 & 0.313 & 0.329 & 0.317 & 0.332 & 0.347 & 0.302 & 0.332 & 0.317 & 0.374 & 0.369 & 0.354 & \textbf{0.491} \\
\cline{2-19}
&Imprv & 87.40\% & 60.98\% & 53.92\% & 69.31\% & 51.08\% & 56.87\% & 49.24\% & 54.89\% & 47.89\% & 41.50\% & 62.58\% & 47.89\% & 54.89\% & 31.28\% & 33.06\% & 38.70\% & --\\
\cline{2-19}
&NDCG & 0.153 & 0.189 & 0.191 & 0.167 & 0.201 & 0.190 & 0.196 & 0.191 & 0.198 & 0.206 & 0.185 & 0.179 & 0.178 & 0.221 & 0.222 & 0.209 & \textbf{0.300} \\
\cline{2-19}
&Imprv & 96.08\% & 58.73\% & 57.07\% & 79.64\% & 49.25\% & 57.89\% & 53.06\% & 57.07\% & 51.52\% & 45.63\% & 62.16\% & 67.60\% & 68.54\% & 35.75\% & 35.14\% & 43.54\% & --\\
\hline
\multirow{4}{*}{IJCAI}
&HR & 0.285 & 0.392 & 0.449 & 0.292 & 0.459 & 0.448 & 0.455 & 0.469 & 0.429 & 0.452 & 0.461 & 0.481 & 0.475 & 0.481 & 0.463 & 0.489 & \textbf{0.532}\\
\cline{2-19}
&Imprv & 86.67\% & 35.71\% & 18.49\% & 82.19\% & 15.90\% & 18.75\% & 16.92\% & 13.43\% & 24.01\% & 17.70\% & 15.40\% & 10.60\% & 12.00\% & 10.60\% & 14.90\% & 8.79\% & --\\
\cline{2-19}
&NDCG & 0.185 & 0.250 & 0.284 & 0.188 & 0.294 & 0.287 & 0.288 & 0.304 & 0.260 & 0.285 & 0.292 & 0.304 & 0.296 & 0.307 & 0.277 & 0.309 & \textbf{0.345}\\
\cline{2-19}
&Imprv & 86.49\% & 38.00\% & 21.48\% & 83.51\% & 17.35\% & 20.21\% & 19.79\% & 13.49\% & 32.69\% & 21.05\% & 18.15\% & 13.49\% & 16.55\% & 12.38\% & 24.55\% & 11.65\% & --\\
\hline
\end{tabular}
\label{tab:target_behavior}
\end{table*}

\begin{table}[h]
	\caption{Rank-based prediction accuracy via varying Top-\textit{N} value measured by \textit{HR@N} and \textit{NDCG@N} on Beibei data.}
	\vspace{-0.05in}
	\centering
    \footnotesize
	\setlength{\tabcolsep}{1mm}
	\begin{tabular}{|c|c|c|c|c|c|c|c|c|}
		\hline
		\multirow{2}{*}{Model}&\multicolumn{2}{c|}{@1}&\multicolumn{2}{c|}{@3}&\multicolumn{2}{c|}{@5}&\multicolumn{2}{c|}{@7} \\
		\cline{2-9}
		&HR&NDCG&HR&NDCG&HR&NDCG&HR&NDCG\\
		\hline
		\hline
		BiasMF & 0.118 & 0.118 & 0.310 & 0.228 & 0.453 & 0.287 & 0.537 & 0.316 \\
		\hline
        NCF-N & 0.123 & 0.123 & 0.317 & 0.232 & 0.447 & 0.283 & 0.530 & 0.315 \\
        \hline
        NGCF-M & 0.167 & 0.167 & 0.376 & 0.286 & 0.496 & 0.337 & 0.569 & 0.361 \\
        \hline
        MBGCN & 0.167 & 0.167 & 0.374 & 0.284 & 0.498 & 0.337 & 0.579 & 0.363 \\
        \hline
        AutoRec & 0.128 & 0.128 & 0.321 & 0.236 & 0.456 & 0.291 & 0.540 & 0.322\\
        \hline
        MATN & 0.184 & 0.184 & 0.361 & 0.286 & 0.467 & 0.330 & 0.543 & 0.356 \\
        \hline
        \emph{\model} & \textbf{0.183} & \textbf{0.183} & \textbf{0.401} & \textbf{0.306} & \textbf{0.527} & \textbf{0.359} & \textbf{0.608} & \textbf{0.389}\\
		\hline
	\end{tabular}
	\vspace{-0.1in}
	\label{tab:vary_k}
\end{table}

\subsection{Model Ablation Study (RQ2)}
To evaluate the rationality of design sub-modules in our \emph{\model} framework, we consider five model variants as follows:

\begin{itemize}[leftmargin=*]

\item \textbf{\emph{w/o LowR}}: We do not integrate the low-rank transformation decomposition in our meta-knowledge learner for encoding behavior heterogeneity. The regular learned meta-transformer $\textbf{V}_{i,k}\in\mathbb{R}^{d\times d}$ is directly used for context embedding. \\\vspace{-0.1in}

\item \textbf{\emph{w/o MFeat}}: The behavior relation learning sub-module is removed in the stage of multi-behavior pattern modeling. \\\vspace{-0.1in}

\item \textbf{\emph{w/o MTask}}: We do not perform the multi-task learning in our meta prediction network with the joint model optimization. \\\vspace{-0.1in}

\item \textbf{\emph{w/o MetaC}}: We do not leverage the designed meta-knowledge learner in encoding the behavior heterogeneous. Instead, we pre-define different transformations for multi-typed behaviors. \\\vspace{-0.1in}

\item \textbf{\emph{w/o MetaP}}: We do not rely on the meta-knowledge learner to learn parameters of our prediction layer. The pre-defined parameter space is adopted for discriminating representation space.

\end{itemize}

\noindent The ablation study results are shown in Table~\ref{tab:module_ablation}. From the evaluation results, we draw the following conclusions:

\begin{itemize}[leftmargin=*]

\item The performance gap between \emph{\model} and the w/o LowR variant indicates the advantage of our designed low-rank transformation decomposition scheme in alleviating the overfitting issue during the heavy transformation operations, by serving as an efficient regularization term in our meta-knowledge learner. \\\vspace{-0.1in}

\item The behavior relation learning shows positive effects for our meta graph neural network during the customized message passing procedure. This observation demonstrates the rationality of our utilized attention layer under multiple representation subspaces, in capturing pair-wise correlations among various interaction behavior types. \\\vspace{-0.1in}

\item The evaluation results shed light on the limitation of single-task learning paradigm without the consideration of cross-type behavior dependency in the prediction phase. This suggests that context behavior relations can not only provide the insightful external knowledge during the explicitly multi-behavior pattern aggregation, but also serve as the supervising signals for model optimization under multi-task learning paradigm. \\\vspace{-0.1in}

\item \emph{\model} outperforms both w/o MetaC and w/o MetaP, due to their simplified design without the incorporation of meta-knowledge learner. The identified meta-knowledge is very important for performing user-specific behavior modeling through a customized meta-learning framework.

\end{itemize}

\begin{table}[t]
    \caption{Ablation study on key components of \model.}
    \vspace{-0.1in}
    \centering
    \small
    \begin{tabular}{c|cc|cc|cc}
        \hline
        Data & \multicolumn{2}{c|}{Beibei} & \multicolumn{2}{c|}{Taobao} & \multicolumn{2}{c}{IJCAI Contest}\\
        \hline
        Metrics & HR & NDCG & HR & NDCG & HR & NDCG\\
        \hline
        \hline
        w/o LowR & 0.6825 & 0.4083 & 0.4385 & 0.2656 & 0.4931 & 0.3195\\
        w/o MFeat & 0.6813 & 0.4049 & 0.3878 & 0.2315 & 0.4865 & 0.3138\\ 
        w/o MTask & 0.6068 & 0.3401 & 0.4577 & 0.2782 & 0.5301 & 0.3392\\ 
        w/o MetaC & 0.6791 & 0.4046 & 0.4647 & 0.2852 & 0.5026 & 0.3199\\ 
        w/o MetaP & 0.6605 & 0.4036 & 0.4868 & 0.2968 & 0.5302 & 0.3399\\ 
        \hline
        \emph{\model} & \textbf{0.6907} & \textbf{0.4103} & \textbf{0.4906} & \textbf{0.2997} & \textbf{0.5319} & \textbf{0.3447}\\
        \hline
    \end{tabular}
    \vspace{-0.1in}
    \label{tab:module_ablation}
\end{table}

\subsection{Effect of Context Behaviors (RQ3)}
\begin{figure}[h]
	\centering
	\subfigure[][Taobao-HR]{
		\centering
		\includegraphics[width=0.395\columnwidth]{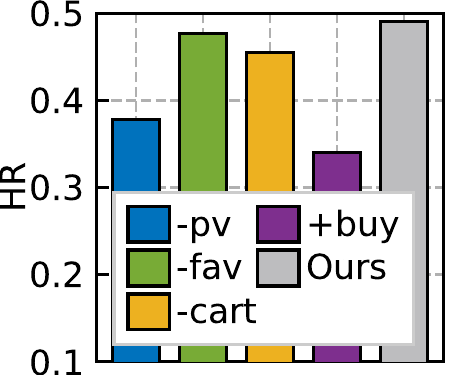}
		\label{fig:ab_tmall_hr}
	}
	\subfigure[][Taobao-NDCG]{
		\centering
		\includegraphics[width=0.39\columnwidth]{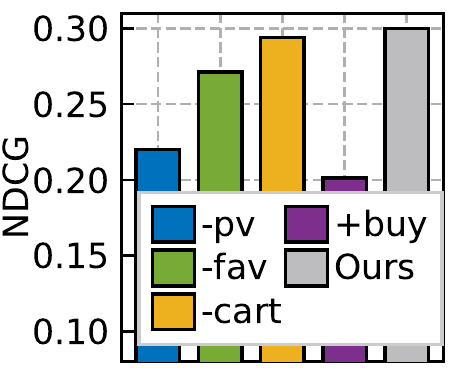}
		\label{fig:ab_tmall_ndcg}
	}
	\subfigure[][IJCAI-HR]{
		\centering
		\includegraphics[width=0.39\columnwidth]{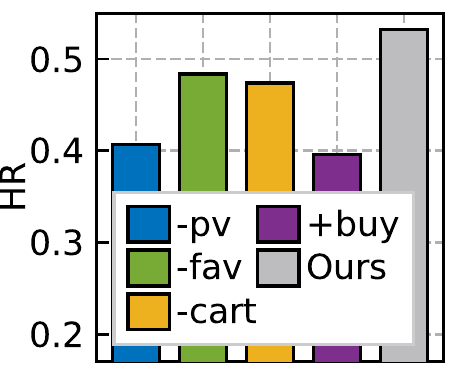}
		\label{fig:ab_ijcai_hr}
	}
	\subfigure[][IJCAI-NDCG]{
		\centering
		\includegraphics[width=0.39\columnwidth]{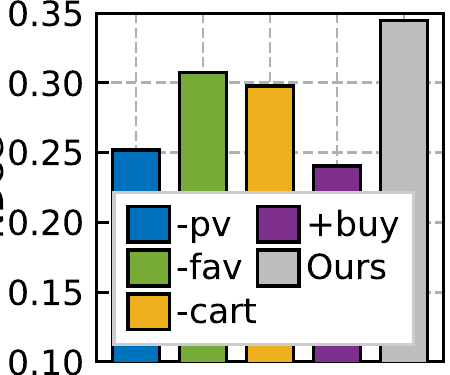}
		\label{fig:ab_ijcai_ndcg}
	}
	\vspace{-0.1in}
	\caption{Impact study of different types of context behavior.}
	\label{fig:beh_ablation}
	\vspace{-0.1in}
\end{figure}

We further perform the ablation study to show the effectiveness of incorporating different context behaviors into our \emph{\model} framework. The evaluation results are presented in Figure~\ref{fig:beh_ablation}. Here, ``-pv'', ``-fav'', ``-cart'' represents our \emph{\model} without the incorporation of individual ``page view'', ``tag-as-favorite'', ``add-to-cart'' behavior, respectively. ``+buy'' is the variant which only relies on the target purchase behaviors themselves to make prediction. From results, we can observe that our framework, which collectively captures user preferences with all types of context behaviors, achieves the best performance as compared to other variants. This demonstrates the effectiveness of learning the interaction heterogeneity to reflect comprehensive multi-behavior patterns.

\subsection{Performance v.s. Sparsity Degrees (RQ4)}

\begin{figure}[h]
	\centering
	\subfigure[][Taobao HR@10]{
		\centering
		\includegraphics[width=0.475\columnwidth]{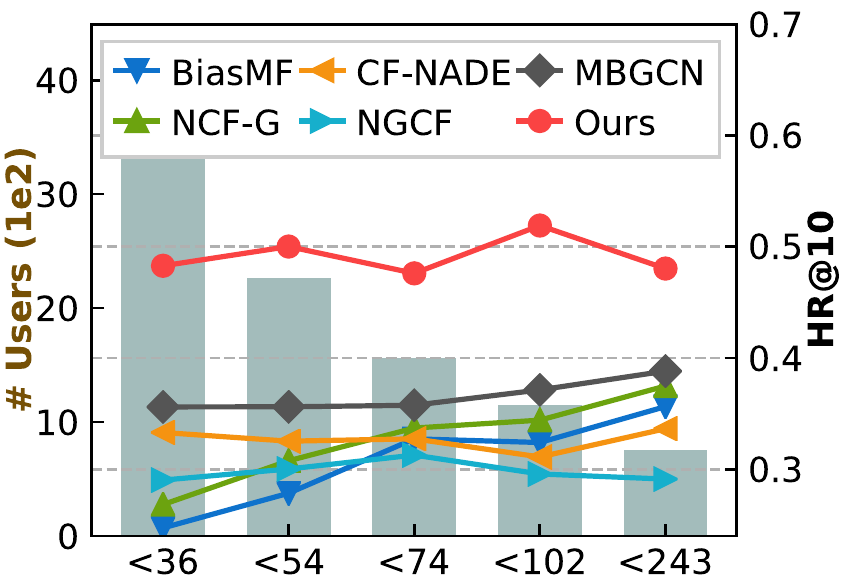}
		\label{fig:sp_tmall_hr}
	}
	\subfigure[][Taobao NDCG@10]{
		\centering
		\includegraphics[width=0.475\columnwidth]{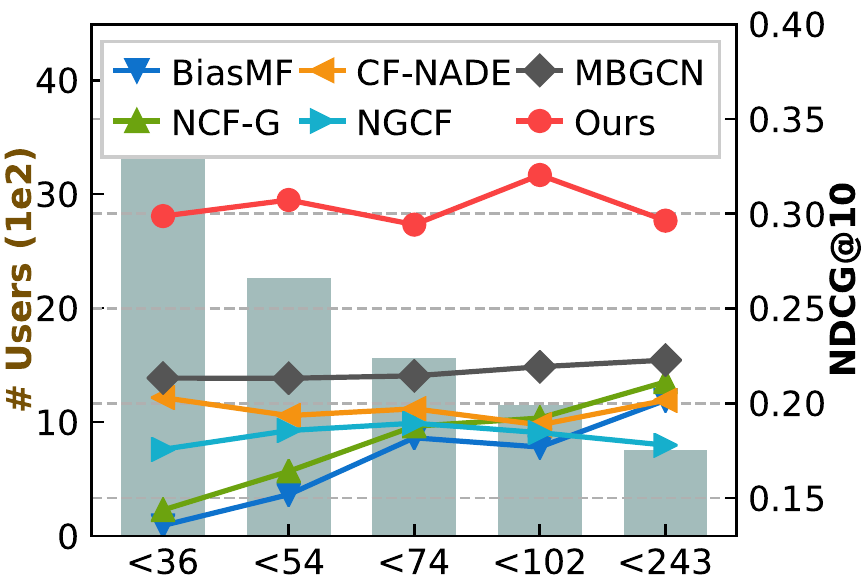}
		\label{fig:sp_tmall_ndcg}
	}
	\vspace{-0.1in}
	\caption{Performance of \model\ and baseline methods \wrt\ different data sparsity degrees on Taobao data.}
	\label{fig:sparsity}
	\vspace{-0.05in}
\end{figure}

Since interaction behavioral patterns vary by users and data sparsity phenomenon is ubiquitous in many real-world recommendation scenarios, we evaluate how is the performance of our new recommendation framework when handling user behavior data with different sparsity degrees. To be specific, x-axis in Figure~\ref{fig:sparsity} represents the number of user interactions, and y-axis indicates the recommendation accuracy (measured by NDCG@10 and HR@10).



From the results, we can observe the consistent performance gain of our \emph{\model} framework over other compared methods, which suggests the model robustness of \emph{\model} in capturing user's preference under different data sparsity degrees. Furthermore, we can notice that it is important to capture the multi-behavior patterns, as \emph{\model} and MBGCN achieve better performance as compared other types of baselines. It is reasonable that incorporating context behaviors (\eg, page view, tag-as-favorite) results in better performance by considering multi-view user preferences. This performance improvement also shows that the incorporation of multi-behaviour information of users can help to alleviate the negative effects of data sparsity in encoding user preference.


\subsection{Hyperparameter Study (RQ5)}
\begin{figure}
    \centering
    \begin{adjustbox}{max width=1.0\linewidth}
    \begin{tikzpicture}
\begin{axis}[
    xlabel={Hidden State Dimensionality $d$},
    ylabel={HR Decrease (\%)},
    xmin=3,xmax=33,
    ymin=-18,ymax=10,
    legend columns=1,
    legend cell align=right,
    grid=both,
    every axis plot/.append style={ultra thick},
    every tick label/.append style={scale=1.3},
    label style={scale=1.8},
    legend style={
        nodes={scale=1.5, transform shape},
        legend image post style={scale=1.5},
        },
    legend style={at={(1,0)},anchor=south east},
    every axis plot post/.append style={
        every mark/.append style={scale=2}
    }
]
\addplot[color={rgb:red,133;green,76;blue,255}, mark=o, mark options={solid}]
table[x=para, y=beibei_hr] {latFactor-buy.txt};
\addplot[color={rgb:red,0;green,157;blue,178}, mark=square, mark options={solid}]
table[x=para, y=tmall_hr] {latFactor-buy.txt};
\addplot[color={rgb:red,245;green,9;blue,11}, mark=triangle, mark options={solid}]
table[x=para, y=ijcai_hr] {latFactor-buy.txt};
\legend{\large Beibei, \large Tmall, \large IJCAI};
\end{axis}
\end{tikzpicture}

\begin{tikzpicture}
\begin{axis}[
    xlabel={Low-Rank Dimensionality $d'$},
    ylabel={HR Decrease (\%)},
    xmin=1,xmax=17,
    ymin=-14,ymax=0.5,
    legend columns=1,
    legend cell align=right,
    grid=both,
    every axis plot/.append style={ultra thick},
    every tick label/.append style={scale=1.3},
    label style={scale=1.8},
    legend style={
        nodes={scale=1.5, transform shape},
        legend image post style={scale=1.5},
        },
    legend style={at={(1,0)},anchor=south east},
    every axis plot post/.append style={
        every mark/.append style={scale=2}
    }
]
\addplot[color={rgb:red,133;green,76;blue,255}, mark=o, mark options={solid}]
table[x=para, y=beibei_hr] {rank.txt};
\addplot[color={rgb:red,0;green,157;blue,178}, mark=square, mark options={solid}]
table[x=para, y=tmall_hr] {rank.txt};
\addplot[color={rgb:red,245;green,9;blue,11}, mark=triangle, mark options={solid}]
table[x=para, y=ijcai_hr] {rank.txt};
\legend{\large Beibei, \large Tmall, \large IJCAI};
\end{axis}
\end{tikzpicture}

\begin{tikzpicture}
\begin{axis}[
    xlabel={Number of GNN Layers $L$},
    ylabel={HR Decrease (\%)},
    xmin=-0.1,xmax=3.1,
    ymin=-30,ymax=8,
    legend columns=1,
    legend cell align=right,
    grid=both,
    every axis plot/.append style={ultra thick},
    every tick label/.append style={scale=1.3},
    label style={scale=1.8},
    legend style={
        nodes={scale=1.5, transform shape},
        legend image post style={scale=1.5},
        },
    legend style={at={(1,0)},anchor=south east},
    every axis plot post/.append style={
        every mark/.append style={scale=2}
    }
]
\addplot[color={rgb:red,133;green,76;blue,255}, mark=o, mark options={solid}]
table[x=para, y=beibei_hr] {gnnlayer.txt};
\addplot[color={rgb:red,0;green,157;blue,178}, mark=square, mark options={solid}]
table[x=para, y=tmall_hr] {gnnlayer.txt};
\addplot[color={rgb:red,245;green,9;blue,11}, mark=triangle, mark options={solid}]
table[x=para, y=ijcai_hr] {gnnlayer.txt};
\legend{\large Beibei, \large Tmall, \large IJCAI};
\end{axis}
\end{tikzpicture}
    \end{adjustbox}
    \begin{adjustbox}{max width=1.0\linewidth}
    \begin{tikzpicture}
\begin{axis}[
    xlabel={Hidden State Dimensionality $d$},
    ylabel={NDCG Decrease (\%)},
    xmin=3,xmax=33,
    ymin=-18,ymax=10,
    legend columns=1,
    legend cell align=right,
    grid=both,
    every axis plot/.append style={ultra thick},
    every tick label/.append style={scale=1.3},
    label style={scale=1.8},
    legend style={
        nodes={scale=1.5, transform shape},
        legend image post style={scale=1.5},
        },
    legend style={at={(1,0)},anchor=south east},
    every axis plot post/.append style={
        every mark/.append style={scale=2}
    }
]
\addplot[color={rgb:red,133;green,76;blue,255}, mark=o, mark options={solid}]
table[x=para, y=beibei_ndcg] {latFactor-buy.txt};
\addplot[color={rgb:red,0;green,157;blue,178}, mark=square, mark options={solid}]
table[x=para, y=tmall_ndcg] {latFactor-buy.txt};
\addplot[color={rgb:red,245;green,9;blue,11}, mark=triangle, mark options={solid}]
table[x=para, y=ijcai_ndcg] {latFactor-buy.txt};
\legend{\large Beibei, \large Tmall, \large IJCAI};
\end{axis}
\end{tikzpicture}

\begin{tikzpicture}
\begin{axis}[
    xlabel={Low-Rank Dimensionality $d'$},
    ylabel={NDCG Decrease (\%)},
    xmin=1,xmax=17,
    ymin=-15,ymax=1,
    legend columns=1,
    legend cell align=right,
    grid=both,
    every axis plot/.append style={ultra thick},
    every tick label/.append style={scale=1.3},
    label style={scale=1.8},
    legend style={
        nodes={scale=1.5, transform shape},
        legend image post style={scale=1.5},
        },
    legend style={at={(1,0)},anchor=south east},
    every axis plot post/.append style={
        every mark/.append style={scale=2}
    }
]
\addplot[color={rgb:red,133;green,76;blue,255}, mark=o, mark options={solid}]
table[x=para, y=beibei_ndcg] {rank.txt};
\addplot[color={rgb:red,0;green,157;blue,178}, mark=square, mark options={solid}]
table[x=para, y=tmall_ndcg] {rank.txt};
\addplot[color={rgb:red,245;green,9;blue,11}, mark=triangle, mark options={solid}]
table[x=para, y=ijcai_ndcg] {rank.txt};
\legend{\large Beibei, \large Tmall, \large IJCAI};
\end{axis}
\end{tikzpicture}

\begin{tikzpicture}
\begin{axis}[
    xlabel={Number of GNN Layers $L$},
    ylabel={NDCG Decrease (\%)},
    xmin=-0.1,xmax=3.1,
    ymin=-30,ymax=8,
    legend columns=1,
    legend cell align=right,
    grid=both,
    every axis plot/.append style={ultra thick},
    every tick label/.append style={scale=1.3},
    label style={scale=1.8},
    legend style={
        nodes={scale=1.5, transform shape},
        legend image post style={scale=1.5},
        },
    legend style={at={(1,0)},anchor=south east},
    every axis plot post/.append style={
        every mark/.append style={scale=2}
    }
]
\addplot[color={rgb:red,133;green,76;blue,255}, mark=o, mark options={solid}]
table[x=para, y=beibei_ndcg] {gnnlayer.txt};
\addplot[color={rgb:red,0;green,157;blue,178}, mark=square, mark options={solid}]
table[x=para, y=tmall_ndcg] {gnnlayer.txt};
\addplot[color={rgb:red,245;green,9;blue,11}, mark=triangle, mark options={solid}]
table[x=para, y=ijcai_ndcg] {gnnlayer.txt};
\legend{\large Beibei, \large Tmall, \large IJCAI};
\end{axis}
\end{tikzpicture}
    \end{adjustbox}
    \vspace{-0.20in}
    \caption{Hyperparameter study of the \model.}
    \vspace{-0.1in}
    \label{fig:hyperparam}
\end{figure}
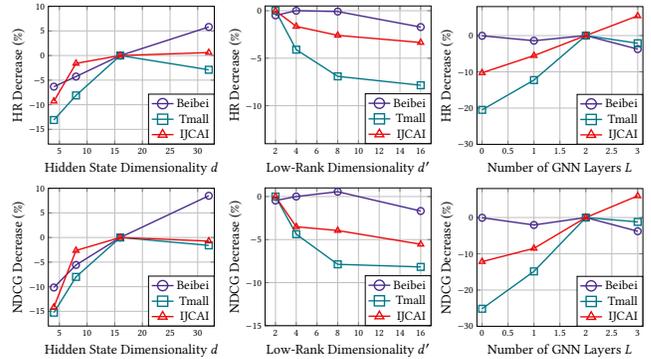

To show the effect of different parameter settings, we perform experiments to evaluate the performance of our developed \emph{\model} framework with different configurations of key hyperparameters, namely, the hidden state dimensionality $d$, low-rank decomposition dimension $d'$, and the depth $L$ of our graph neural network. We show the results in Figure~\ref{fig:hyperparam}. Considering that accuracy vary by different datasets in terms of HR and NDCG metric ranges, the y-axis of Figure~\ref{fig:hyperparam} represents the estimated performance decrease/increase percentage of each parameter-specific performance as compared to the best performance with the optimal hyperparameter settings. The discussions are summarized as follows. 

\begin{itemize}[leftmargin=*]

\item \textbf{Hidden State Dimensionality: $d$}. In our framework, the latent state dimensionality $d$ is searched from 4 to 32. We can observe that with the increasing of embedding dimension from 4 to 16, the recommendation performance improves due to the stronger representation feature space. However, a larger embedding dimensionality does not always bring stronger model representation ability in learning multi-behavioral relations for recommendations. This is caused by the model overftting. \\\vspace{-0.1in}

\item \textbf{Low-rank Decomposition Dimension $d'$}. We can notice that smaller value of $d'$ in low-rank transformation decomposition is more beneficial for learning meta-knowledge from behavior heterogeneity. The reason lies in that the relatively small value of $d'\leq 4$ could reduce the computational cost and improve the model robustness. This observation again confirms the rationality of our design low-rank decomposition scheme in learning meta-knowledge from multi-behavior user data. \\\vspace{-0.1in}

\item \textbf{Depth of Graph Neural Paradigm $L$}. By stacking two graph neural layers, \emph{\model}-2 performs better than \emph{\model}-1, which suggests the positive effect of high-order connectivity injection. However, with the increasing of depth of graph model, the performance starts to deteriorate. This is because stacking more embedding propagation layers may involve noise signals for modeling collaborative effects between users and items, which leads to the over-smoothing~\cite{chen2020measuring}.

\end{itemize}

\subsection{Model Interpretation Study (RQ6)}
\begin{figure}
    \centering
    \includegraphics[width=0.99\columnwidth]{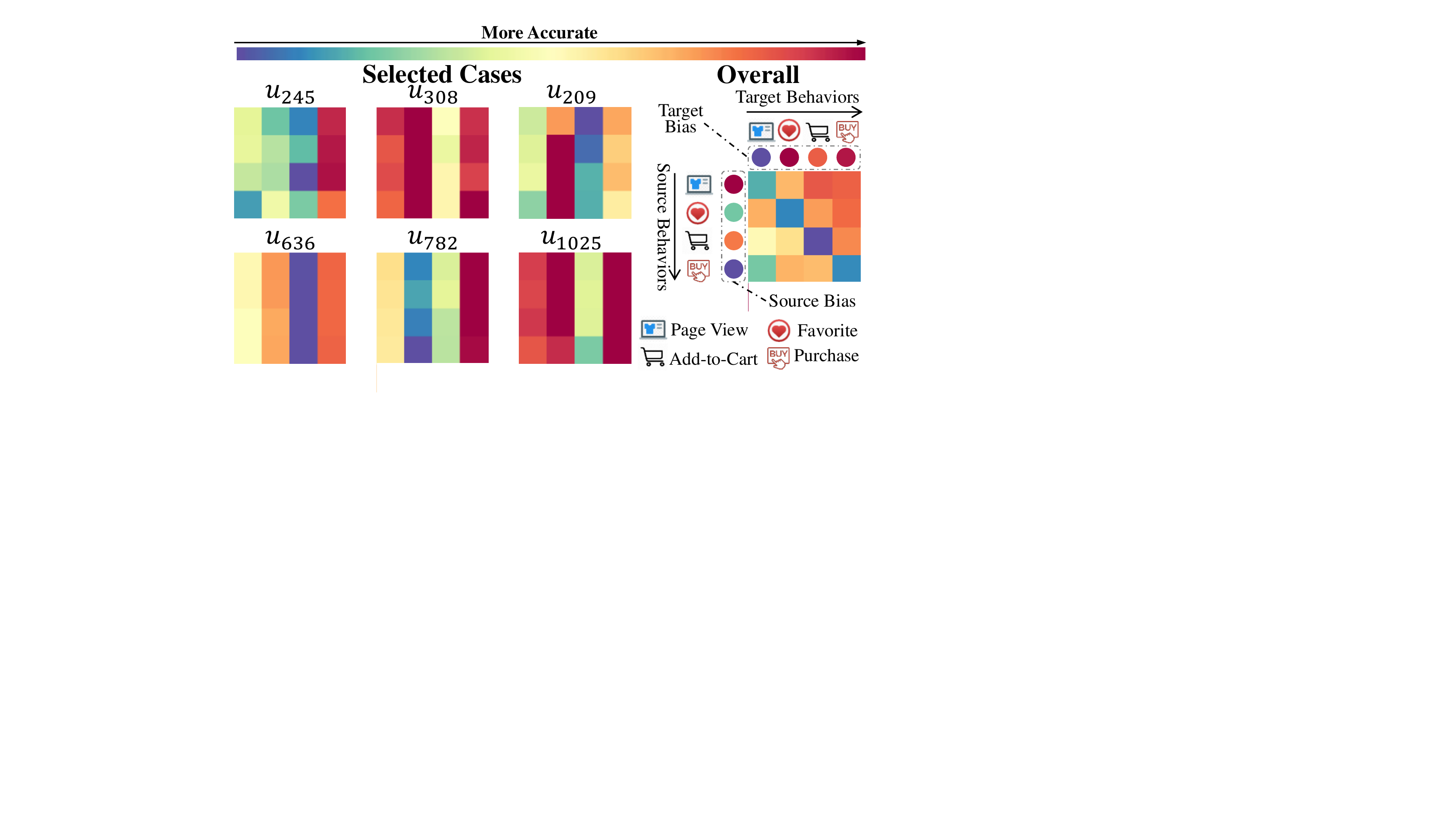}
    \vspace{-0.05in}
    \caption{Case study on the explainability of \emph{\model}. Source bias represents the helpfulness of individual behavior in serving as context behavior in providing auxiliary information. Target bias represents the usefulness for predicting this target behavior based on the context behavior representation. The type-wise behavior dependency is reflected with heatmap matrix, \ie, six heatmap $4\times 4$ matrices corresponding to six extracted users, one heatmap $4\times 4$ matrix presents the overall dependencies across different types of behaviors from all users.}
    \label{fig:case_study}
    \vspace{-0.15in}
\end{figure}

We further perform case studies with sampled case examples to show the interpretation capability of \emph{\model} by capturing the cross-type behavior dependent structures. In the experiments, we visualize the loss value during our training phase after the model convergence, so as to reflect the helpfulness of individual context behavior in predicting the target behavior. In our joint learning framework, each type of user-item interaction behavior could be considered as context source behavior and target behavior. In such cases, we generate a $4\times 4$ weight matrix from six sampled users (\eg, $u_{245}$, $u_{308}$) corresponding to page view, tag-as-favorite, add-to-cart and purchase (as shown in Figure~\ref{fig:case_study}). In general, we can observe that page view behaviors contribute more on predicting other types of behaviors in most cases. The overall helpfulness evaluation results across users are also illustrated with the heatmap matrix on the right side. Similar results can be observed. The potential reason for this observation lies in the larger amount of page view activities of users compared with other types of user interactions. General speaking, all above observations could show the interpretation ability of our \emph{\model} architecture in capturing complex dependency across different types of user behaviors and characterize the underlying helpfulness of context behavior during the prediction.

\section{Related Work}
\label{sec:relate}

\subsection{Collaborative Filtering With Deep Learning}
In recent years, augmenting collaborative filtering techniques with deep neural networks has become increasingly popular, due to the strong capability of neural architectures in learning complex interactive patterns between users and items~\cite{chen2019joint,yao2020efficient}. There exist various types of deep learning paradigms, with structures suited to collaborative filtering architectures. Specifically, NCF~\cite{he2017neuralncf} proposed to enhance the feature interaction between projected user and item embeddings with Multi-Layer Perceptron component, so as to address the limitation of inner-product operation in the recommendation phase. DMF~\cite{xue2017deep} extended the matrix factorization with neural networks to map users and items into a common latent space. Autoencoder and its model variants have also been utilized for collaborative filtering tasks with the reconstruction-based encoder-decoder learning over user-item interactions~\cite{sedhain2015autorec,wu2016collaborative}. Different from the above deep collaborative filtering models which merely consider single-typed interactive behavior of users, our \model\ explores the users' multi-behavioral interactions and effectively incorporates the cross-behavior inter-dependent knowledge into the recommender systems, to boost the performance.

\subsection{Multi-Relation Recommendation Models}
Many methods were proposed to enhance the representation capability of recommender systems by characterizing various relations from the side information of users and items~\cite{liu2020heterogeneous,heterogeneousrecsys21}. Let us discuss some recent literature: a research line of recommendation models proposed to incorporate online social relations of users to assist the user representation, by considering the influence between different users~\cite{wu2019neural,social2021knowledge,chen2019efficient}. Furthermore, knowledge graph information has also become an important data source to enhance existing recommendation frameworks. These knowledge-aware recommendation models made use of connections between different entities to supplement the interaction learning between users and items~\cite{wang2019knowledge,wang2019kgat,zhu2020knowledge}. For instance, a knowledge-aware attentive reasoning scheme is proposed to aggregate information from user behaviors and relational paths in the generated knowledge graphs~\cite{zhu2020knowledge}.


\subsection{Graph Neural Networks for Recommendation}
Recent years have witnessed the effectiveness of graph neural networks (GNNs) for mining various relational data~\cite{chami2019hyperbolic,2019heterogeneous}. In general, the key idea of GNN models is to perform graph-structured learning with message passing architectures for aggregating feature information from neighborhood nodes. For example, Graph Convolutional Network (GCN)~\cite{gao2018large}, GraphSAGE~\cite{hamilton2017inductive} and Graph Attention Network (GAT)~\cite{wang2019heterogeneous} are several representative GNN models which aggregate feature information from neighboring vertices, by utilizing convolutional operation, or attentional mechanism. GNNs have also been utilized to augment the user-item interaction modeling based on graph structures. For example, NGCF~\cite{wang2019neural} and ST-GCN~\cite{zhang2019star} are message passing recommendation methods which design aggregation encoder with graph convolutional network. Zheng~\etal~\cite{zheng2020price} learns price-aware purchase intent of users with a GCN-based architecture. Inspired by the above research work, we propose a new neural approach \model\ within the broader graph structure for multi-behavior recommendation, to tackle the technical challenges for dealing with the interaction diversity.

\section{Conclusion}
\label{sec:conclusion}

This work develops a novel multi-behavior enhanced recommendation framework with graph meta network for learning interaction heterogeneity and diversity. Our framework explicitly incorporates the inter-correlations of multiple types of user behavior into the collaborative filtering architecture, which are usually neglected in most existing recommendation methods. In addition, our proposed framework incorporates the meta-knowledge network into a meta-learning paradigm, which endows the multi-behavior pattern modeling in a customized manner. Extensive experiments on three large-scale E-commerce datasets verify the effectiveness of \model. Several promising direction of future work are: First, we plan to extend the developed \model\ framework by incorporating side context information into the modeling process of user's preference, such as user's profile information and items' text descriptions. Second, a time-sensitive model is deserved to be investigated in order to handle the new arriving data streaming of users' behavior in a timely manner, which benefits the real-time recommendation.

\section*{Acknowledgments}
We thank the reviewers for their valuable comments. This work is supported by National Nature Science Foundation of China (62072188), Major Project of National Social Science Foundation of China (18ZDA062), Science and Technology Program of Guangdong Province (2019A050510010).

\clearpage

\bibliographystyle{ACM-Reference-Format}
\balance
\bibliography{sigproc}

\end{document}